\documentclass[jcp,amsmath,amssymb,reprint]{revtex4-1}

\usepackage{color}
\usepackage{graphicx}
\usepackage{dcolumn}
\usepackage{bm}

\begin{document}

\preprint{}

\title{Structure of benzothiadiazine at zwitterionic phospholipid cell membranes}

\author{Zheyao Hu,  Jordi Mart\'\i$^{\star}$}%
\email{zheyao.hu@upc.edu, jordi.marti@upc.edu}
\altaffiliation{Department of Physics, Technical University of Catalonia-Barcelona Tech, 
B5-209 Northern Campus UPC, 08034 Barcelona, Catalonia, Spain$\;$ (Corresponding author)}
\author{Huixia Lu}
\email{huixia.lu@sjtu.edu.cn}
\altaffiliation{School of Pharmacy, Shanghai Jiaotong University, Shanghai,China}

\date{\today}

\begin{abstract}
The use of drugs derived from benzothiadiazine,  which is a bicyclic heterocyclic benzene derivative,  has become a widespread treatment for diseases such as hypertension (treated with diuretics such as bendroflumethiazide or chlorothiazide), low blood sugar (treated with non-diuretic diazoxide) or the human immunodeficiency virus, among others.  In this work we have investigated the interactions of benzothiadiazine with the basic components of cell membranes and solvents such as phospholipids,  cholesterol,  ions and water.  The analysis of the mutual microscopic interactions is of central importance to elucidate the local structure of benzothiadiazine as well as the mechanisms responsible for the access of benzothiadiazine to the interior of the cell.  We have performed molecular dynamics simulations of benzothiadiazine embedded in three different model zwitterionic bilayer membranes made by dimyristoilphosphatidylcholine,  dioleoylphosphatidylcholine,  1,2-dioleoyl-sn-glycero-3-phosphoserine and cholesterol inside aqueous sodium-chloride solution in order to systematically examine microscopic interactions of benzothiadiazine with the cell membrane at liquid-crystalline phase conditions.  From data obtained through radial distribution functions, hydrogen-bonding lengths
and potentials of mean force based on reversible work calculations, we have observed that benzothiadiazine has a strong
affinity to stay at the cell membrane interface although it can be fully solvated by water in short periods of time.  Furthermore, 
benzothiadiazine is able to bind lipids and cholesterol chains by means of single and double hydrogen-bonds of different characteristic lengths.
\end{abstract}

\maketitle

\section{\label{intro}Introduction}

Cell membranes are responsible of the separation and exchange of cell contents from external environments.  Membranes only allow 
the movement of compounds generally with small molecular size (within the nanometric range) in or out the cell.  For such a reason, 
the knowledge of the mechanisms responsible for the exchange of small peptides and drugs inside the membrane are of greatest 
scientific interest. The principal components of human cellular membranes are phospholipids, cholesterol and proteins, always 
inside electrolyte aqueous solution. Phospholipid membranes consist of two leaflets of amphiphilic lipids with a hydrophilic head 
and one or two hydrophobic tails which self-assemble due to the hydrophobic effect\cite{nagle2000structure, mouritsen2005life}. 
Among a wide variety of lipids, dimyristoilphosphatidylcholine (DMPC,  $C_{36}H_{72}NO_{8}P$) are saturated phospholipids incorporating a choline as a headgroup and a tailgroup formed by two myristoyl chains. They are usually synthesised to be used for research purposes (studies of liposomes and bilayer membranes). Their properties are very similar to those of dipalmitoylphosphatidylcholine (DPPC, $C_{40}H_{80}NO_{8}P$) which has the same structure but slightly longer tails, being 
a major constituent of pulmonary surfactants of lungs (about 40\%).  Differently,  1,2-dioleoyl-sn-glycero-3-phosphocholine 
(DOPC, $C_{44}H_{84}NO_{8}P$) and 1,2-dioleoyl-sn-glycero-3- phospho-L-serine (DOPS,  $C_{42}H_{77}NO_{10}P$) are unsaturated species very common in the tissues forming the most essential human organs.  Cholesterol ($C_{27}H_{46}O$), is 
a sterol playing a central role in maintaining the structure of the membrane and regulating their functions\cite{ohvo2002cholesterol,mcmullen2004cholesterol}. It induces the membrane to adopt a liquid-ordered phase with positional disorder and high lateral mobility\cite{mouritsen2005life} and also regulates the fluidity of the membrane.  

There exists a big variety of experimental techniques useful to explore membrane organisation and molecular interactions of small 
probes within the membrane, such as: NMR, neutron diffraction, X-ray scattering or IR,  Raman and fluorescence spectroscopy\cite{petrache2000area,sanchez2011tryptophan,leyton2012experimental,liu2013fluorescence}.  Among the latest
 techniques, fluorescence-lifetime microscopy\cite{digman2008phasor} can be combined with spectral information to report 
basic information of aspects such as metabolic profiles, photophysics or dipolar relaxations\cite{malacrida2017multidimensional}. 
However, the direct information on atomic interactions and local structures can only by accessed by computer simulations at 
the all-atom level, as it is the aim of the present work.

Heterocyclic molecules play an important role in medicine and are closely related to human 
life\cite{horton2003combinatorial,sharma2017heterocyclic,capdeville2002glivec}, agriculture\cite{matheron2004activity} 
and industry\cite{corbet2006selected},  as the main fields of application.  Such compounds are common parts of commercial 
drugs having multiple applications based on the control of lipophilicity, polarity and molecular hydrogen bonding capacity. 
Among them,  3,4-dihydro-1,2,4-benzothiadiazine-1,1-dioxide ($C_{7}H_{8}N _{2}O_{2}S$,  DBD,  benzothiadiazine from now on) 
is a bicyclic heterocyclic benzene derivative containing two nitrogen atoms and one sulphur within the heterocyclic group. Benzothiadiazine derivatives have wide pharmacological applications, such as  diuretic\cite{platts1959hydrochlorothiazide,taylor1963pharmacology}, antiviral\cite{martinez1999novel,martinez2001anti,das2011recent,m20123d},  anti-inflammatory\cite{tait2005synthesis}, anticancer\cite{kamal2008synthesis,ma2019design}or the regulation of the central nervous system\cite{norholm2013synthesis,larsen2016synthesis} to mention a few.  Because of its importance as the main common structure of the DBD family, in this work we have selected the 3,4-dihydro-1,2,4-benzothiadiazine-1,1-dioxide species as our target to
explore its affinity to the cell membrane and its local structure and hydrogen-bonding characteristics. Up to the best 
of our knowledge,  no microscopical studies of DBD at the interface of phospholipid membranes are yet available.  
So the main goal of the present study is to report structural and energetic results obtained from all-atom
computer simulations of three classes of membranes for the first time.

In this paper we provide the details of the simulations in section \ref{met} and explain the main results of the work in 
section \ref{res}, focusing our attention especially on the local structures of DBD (section \ref{struc}) and also on the 
free-energy barriers of DBD and specific atomic sites, described in section \ref{pmfs}. Finally, some concluding remarks 
are outlined in section \ref{concl}.\\

\section{Methods}
\label{met}

\begin{figure}[ht]
\begin{center}
\includegraphics[scale=0.7,angle=0]{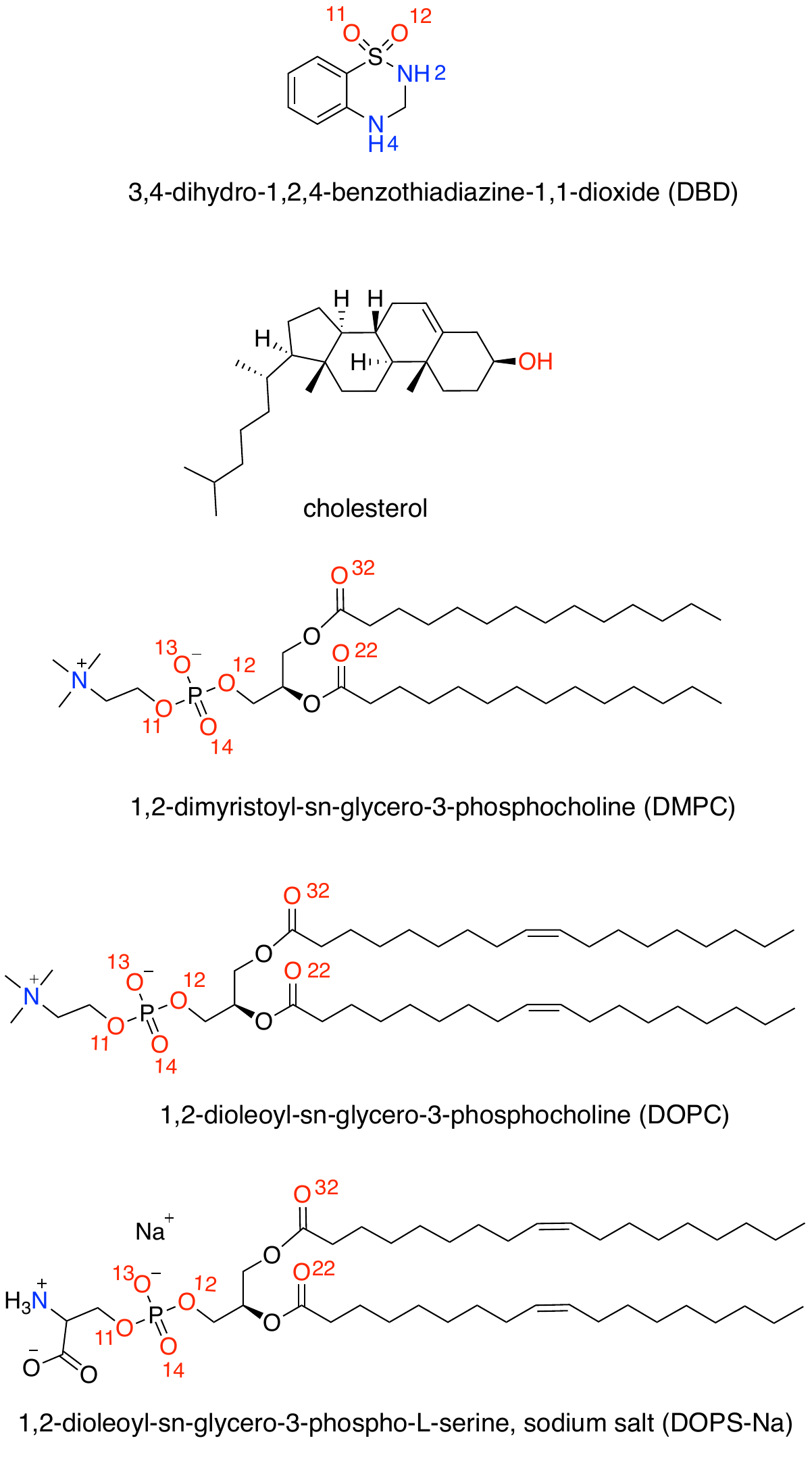}
\caption{\label{fig1}Sketches of the backbone structures of DBD, DMPC,  DOPC,  DOPS and cholesterol. Hydrogens bound to carbon not 
shown. The highlighted sites of each species will be referred in the text by the labels defined here. }
\end{center}
\end{figure}

Three models of lipid bilayer membranes in an aqueous solution have been constructed using the CHARMM-GUI 
web-based tool\cite{Jo2008a,Jo2009}. The membrane components and the amount of particles of each class are
summarised in Table~\ref{tab1}. The lipids have been distributed in two leaflets embedded inside the electrolyte solution.
We have considered three different setups, namely: (1) a bilayer membrane formed only by DMPC, that we will label as 'system 1'; 
(2) a membrane formed by neutral DOPC and DOPS,  labelled as 'system 2' and finally (3) a membrane made by neutral DOPC and DOPS associated to Na$^+$, together with cholesterol in ionic solution, labelled as 'system 3'.  System 1 is the simplest prototype cell 
membrane design, whereas system 3 is the most reallistic, complete one of the three,  including electrolyte sodium-chloride
solution at 0.15 M concentration, typical of human body.  System 2 is a model with two clases of lipids, a kind of intermediate 
setup between the other two.  Sketches of all species are reported in Fig.~\ref{fig1}.  Each phospholipid was described 
with atomic resolution (DMPC had 118 sites, DOPC 138 sites, DOPS 131 sites and cholesterol 74 sites). In all simulations, 
a single prototype drug, benzothiadiazine (20 sites, also represented in Fig.~\ref{fig1}) has been chosen for the evaluation 
of its interactions with the main components of the membrane. Water has been represented by rigid TIP3P\cite{Jorgensen1983} molecules.  In all cases, we run the simulations at the fixed pressure of 1 atm and at the temperature of 303.15 K, well 
above the crossover temperatures for pure DMPC,  DOPC and DOPS needed to be at the crystal liquid phase (295, 
253 and 262 K, respectively)\cite{chen2018determination}).

\begin{table}
\caption{\label{tab1} Characteristics of the systems simulated in this work. 10000 water molecules considered in all cases.}
\fontsize{7}{10}\selectfont
\begin{tabular} {|c|c|c|}  \hline
System & Lipids & ions \\      \hline
   1 &  200 DMPC &  0  \\
   2 & 160 DOPC,  40 DOPS  & 0     \\
   3  & 112 DOPC,  28 DOPS,  60 Cholesterol  &  55 Na$^+$,  27 Cl $^-$   \\  
    \hline
\end{tabular}
\end{table}

Molecular dynamics (MD) simulations were performed by means of the NAMD2 simulation package\cite{Phillips2005}.
In all cases, the temperature was controlled by a Langevin thermostat\cite{berendsen1984molecular} with a damping coefficient of 1 ps$^{-1}$, whereas the pressure was controlled by a Nos\'{e}-Hoover Langevin barostat\cite{Feller1995} with a damping time of 50 fs.  
In the isobaric-isothermal ensemble,  i.e.  at the condition of constant number of particles (N), pressure (P) and temperature (T),  equilibration periods for all simulations were of more than 50 ns.  In all cases, we recorded statistically meaningful trajectories of 
more than 200 ns through several production runs. The simulation boxes had different sizes because of the composition of the membrane.  For instance, in system 1, the size was of 77.9$\;$\AA $\times$ 77.9$\;$\AA $\times$ 85.1$\;$\AA.  For system 2
the size was of 87.1$\;$\AA $\times$ 87.1$\;$\AA $\times$ 72.9$\;$\AA, whereas for system 3 it was of 71.3$\;$\AA $\times$ 
71.3$\;$\AA $\times$ 101.7$\;$\AA.  

We have considered periodic boundary conditions in the three directions of space. The simulation time step was fixed to 2 fs.  The  CHARMM36 force field\cite{Klauda2010,Lim2012} was adopted for lipid-lipid and lipid-protein interactions.  We selected the
version CHARMM36m\cite{huang2013charmm36},  able to reproduce the area per lipid for the most relevant phospholipid 
membranes, in excellent agreement with experimental data. All bonds involving hydrogens were set to fixed lengths, allowing fluctuations of bond distances and angles for the remaining atoms.  Van der Waals interactions were cut off at 12 \AA$\;$ with a smooth switching function starting at 10 \AA.  Finally,  long ranged electrostatic forces were computed using the particle mesh Ewald method\cite{Essmann1995}, with a grid space of about 1 \AA,  updating such electrostatic interactions every time step of each simulation.
  
\section{Results and discussion}
\label{res}

The three classes of bilayer phospholipid membranes considered in this work were previously simulated and their main
characteristics were thoroughly analysed (see for instance Ref. \cite{yang2014diffusion} for DMPC and Ref.\cite{lu2020long} for DOPC-DOPS),  finding suitable values of the area per lipid  $A$ (by means of the deuterium order parameter) and thickness of the membranes (defined by the average distance between phosphorus of both leaflets of the membrane), in good agreement with available experimental and simulation data.  For the calculation of $A$ we considered the total surface along the $XY$ 
plane (plane parallel to the bilayer surface) divided by the number of lipids (eventually plus cholesterol) in one lamellar 
layer\cite{pandey2011headgroup}.

\subsection{Local structure of benzothiadiazine}
\label{struc}

\subsubsection{Radial distribution functions}

The local structure of a multicomponent system is usually analysed by means of the (normalised) atomic radial distribution 
functions (RDF) $g_{AB}(r)$.  For a species $B$ close to a tagged species $A$, they are given by

\begin{equation}
g_{AB}(r) = \frac{V\,\langle n_{B}(r)\rangle}{4\,N_{B}\,\pi r^2\;\Delta r},
\label{eq1}
\end{equation}

where $n_{B}(r)$ is the number of atoms of species $B$ surrounding a given atom of species $A$ inside a spherical 
shell of width $\Delta r$. $V$ stands for the total volume of the system and $N_B$ is the total number of particles of species 
$B$.  $g_{AB}(r)$ physical meaning stands for the probability of finding a particle $B$ at a given distance $r$ of a particle $A$.
With the usual normalisation taken here,  RDF will tend to 1 at long distances, when reaching the average density of the system.

\begin{figure}[ht]
\begin{center}
\includegraphics[scale=0.4]{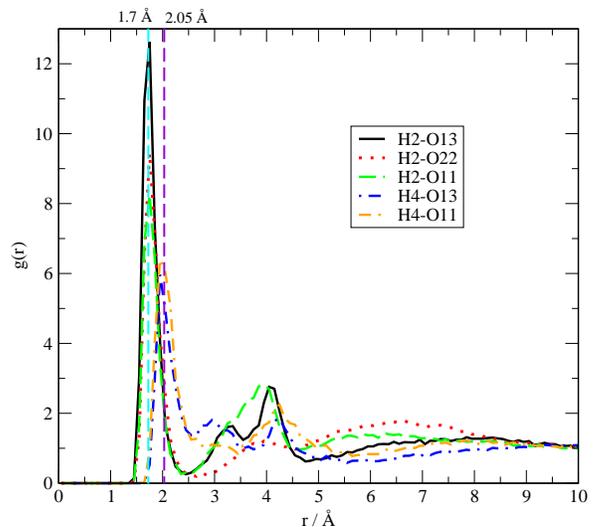}
\caption{\label{fig2}Radial distribution functions for DBD with DMPC (system 1).}
\end{center}
\end{figure}

We have evaluated the local structure of the DBD molecule solvated by lipids and cholesterol.  Among the myriad of 
possible RDF that we can compute,  we will restrict ourselves to report a selection of the most relevant ones.  The results 
are presented in Figures ~\ref{fig2}, ~\ref{fig4} and ~\ref{fig5}, corresponding to the simulated systems 1,  2 and 3 as 
described in Table~\ref{tab1}.  All RDF show some fluctuations in their profiles due to statistical noise.  As a general feature,  
the active sites of DBD capable of forming hydrogen bonds (HB) are hydrogens 'H2' and 'H4' and oxygens 
'O11' and 'O12' (see Fig.~\ref{fig1}).  We will see below that the role of such atomic sites is not unique,  participating
in a wide variety of HB of different length, ranging between 1.7 and 2.1~\AA.  We can observe a clear first coordination 
shell in all cases associated to the binding of DBD to lipid and cholesterol species (and eventually water as well in the 
case of system 3) together with lower second shells centred around 4~\AA.  

\begin{figure}[ht]
\begin{center}
\includegraphics[scale=0.28]{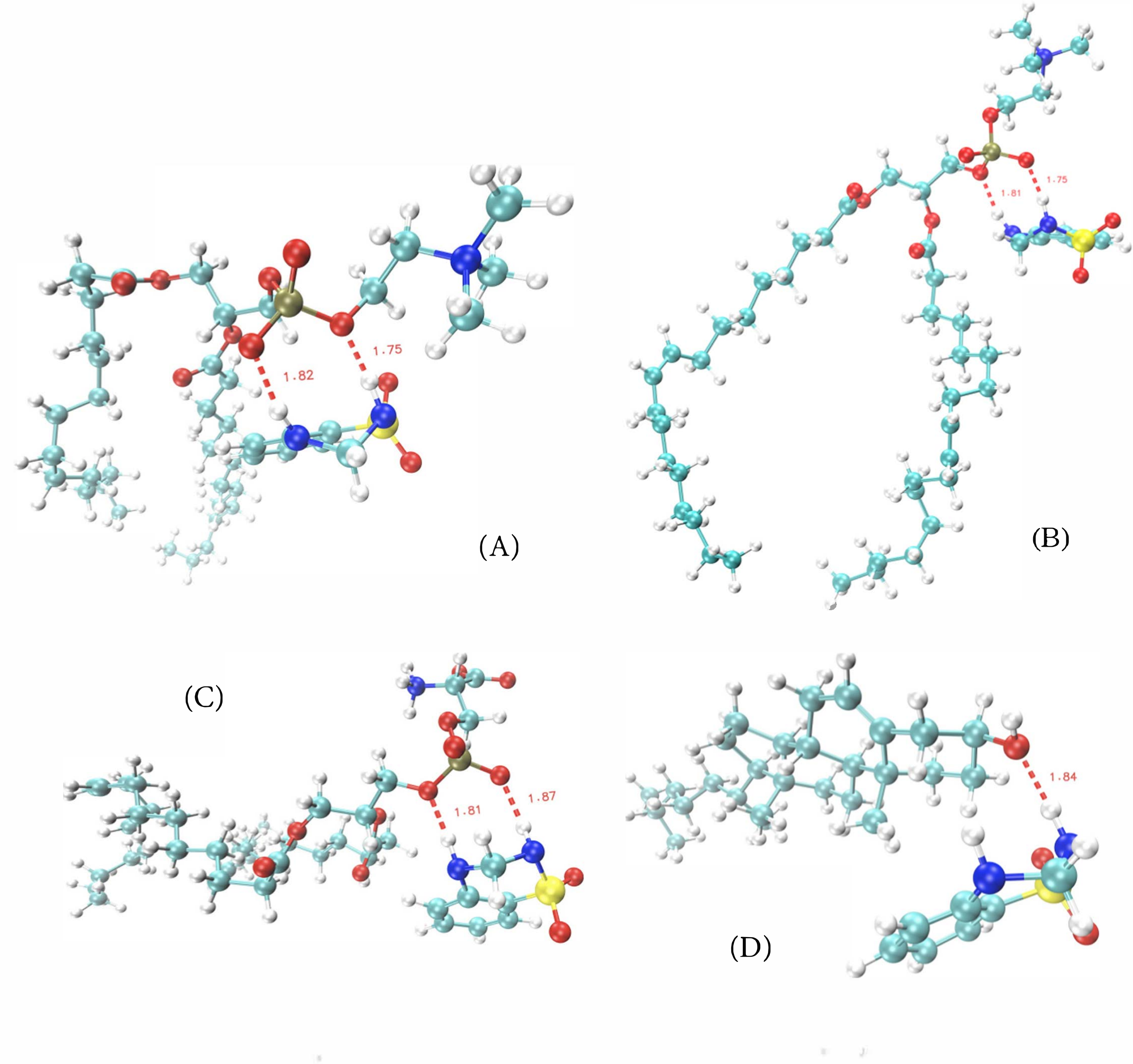}
\caption{\label{fig3}Snapshots of typical DBD-lipid and DBD-cholesterol bonds.  Hydrogen bonds depicted in red, oxygens in red, nitrogen in blue, carbon in cyan, phosphorus in brown and sulphur in yellow. (A) DBD-DMPC, (B) DBD-DOPC, (C) DBD-DOPS, (D) DBD-cholesterol.}
\end{center}
\end{figure}

\begin{figure}[ht]
\begin{center}
\includegraphics[scale=0.4]{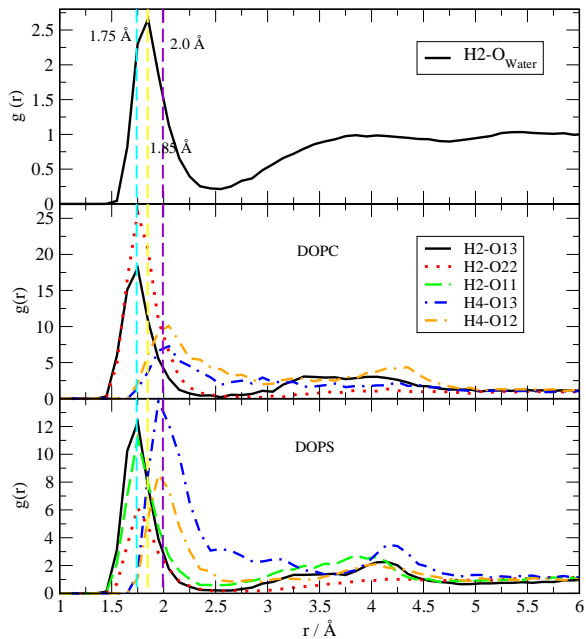}
\caption{\label{fig4}Radial distribution functions for DBD with water, DOPC and DOPS (system 2).}
\end{center}
\end{figure}

\begin{figure}[ht]
\begin{center}
\includegraphics[scale=0.5]{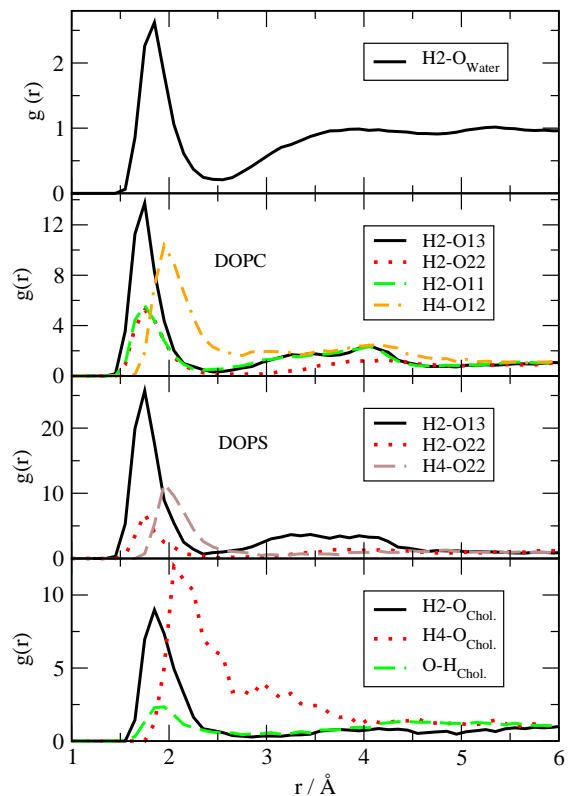}
\caption{\label{fig5}Radial distribution functions for DBD with water, DOPC,  DOPS and cholesterol(system 3).}
\end{center}
\end{figure}

The structure of DBD in system 1 is the simplest, as expected.  We can distinguish DBD's hydrogens H2 and H4 both forming 
HB with DMPC.  H2 is able to bind oxygens 'O13', 'O22' and 'O11' of DMPC, located in different regions of the lipid structure (see Fig.~\ref{fig1}).  In many cases, the HB length (location of first maximum of the RDF) is a short distance of 1.7~\AA,  typical of the 
binding of small-molecules to cell membranes, such as tryptophan to dipalmytoilphosphatidylcholine (DPPC,  see for instance 
the review \cite{marti2021microscopic}). It should be pointed out that using fluorescence spectroscopy,  Liu et 
al.\cite{liu2013fluorescence} obtained values for the hydrogen-bond lengths of tryptophan-water between 1.6 and 
2.1~\AA,  i.e of the same range than those reported here. The largest peak (see Fig.~\ref{fig2}) corresponds to H2 forming 
HB with oxygens of the phosphoryl group O13 and O14,  since both oxygen sites are sharing a negative charge and their 
contributions have been averaged and labelled 'O13', for the sake of simplicity. The same rule has been applied to oxygens 
'O22' and 'O32', although labels 'O11' and 'O12' have been analysed separately,  given their different location in the lipid 
structure.  This way of grouping equivalent lipid sites has been applied to all lipid species (DMPC, DOPC and DOPS).  Concerning 
hydrogens H4 of DBD, they can be either connected to O13 or O11, but not to O22.  Interestingly, in the case of DBD's H4, 
HB lengths are in the range of 2-2.1~\AA, significantly longer than those formed by H2 (range around 1.6 to 1.8~\AA).
We have analysed the frequency of such HB and observed that they tend to occur in simultaneous pair structures, as it is 
sketched in Fig.~\ref{fig3}. So,  in plots (A), (B) and (C) we can observe that DBD's H2 hydrogen-bonded to DMPC's O13-O14 and DBD's H4 bonded to DMPC's O11 happen simultaneously.  Further, the angles formed by O-H-N (oxygen from lipids and H-N 
from DBD) are essentially flat,  with values around 180$^\circ\pm 30^\circ\;$ in a rough estimation.  We observed such 
double bindings for periods of time longer than 5 ns (considering statistical average along the 200 ns long MD trajectory).  
The group of eight atomic sites involved in such double hydrogen-bond, closed-ring structures reassembles the 
special structure observed in some oncogenic proteins (KRas-4B) where salt bridges were found as the main 
responsible of the anchoring mechanism of the protein in the cell membrane\cite{lu2020long}.  This fact is 
qualitatively different of the single HB formed between small-molecules and lipid bilayer membranes observed 
in previous works (see Refs.\cite{lu2018effects, lu2019binding} for instance) where the small-molecule (melatonin 
or tryptophan) was able to bridge several lipid units and keep them together for several ns, probably due to the 
relative large size of the small-molecule and to the situation of their binding sites quite far away each other, differently 
of the very close binding sites of DBD.

Similar findings hold for system 2, where DBD is located at the interface of a DOPC-DOPS bilayer membrane.  According to the
results reported in Figure~\ref{fig4},  we have distinguished three RDF for DBD-water, DBD-DOPC and DBD-DOPS.  In this setup,
DBD is able to be solvated by water for periods of the order of 20 ns. In such a case, we have located the corresponding HB
formed by DBD's H2 and oxygens of water, with the characteristic signature length of the water's HB of 1.85~\AA 
(see \cite{marti1999analysis,modig2003temperature}). We have not found significant evidence of H4-water oxygen 
hydrogen-bonding.  Regarding DBD-lipid bindings, we have observed that DBD’s H2 is more likely to establish HB with 
DOPC’s O13 and O22, unlikely with O11 or O12. When analysing DOPS, H2 is also able to hydrogen-bond to O13 but not to 
O22 and,  conversely,  it can produce stable HB with O11. So the interactions of DBD are slightly different regarding DOPC 
or DOPS. However, the binding of DBD's H4 are essentially the same for the two lipid species: it can be bound to O13 and 
O11 in both cases.  The signature HB distances have been determined to be almost the same as in the DMPC case: 
1.75~\AA$\;$  for H2 bindings and 2.0~\AA$\;$ for the H4 ones.  All secondary shells have been also found centred 
around 4~\AA .  In system 2 we have also detected double HB between DBD and the two lipid classes (see Fig.~\ref{fig3}, 
snapshots (B) and (C) where DOPC and DOPS, respectively are shown): H2 with O13 together with H4 with O12.

Finally,  the most complete and more realistic setup with system 3 has revealed several RDF similar to system 2, but some
new features: (1) DBD's H2 is able to establish some HB with DOPC's O11,  but with maxima less marked than H2-O13; (2)
H2 is only bound to O13 of DOPS and (3) H4 connections are less and weaker than in system 2. We believe that this general
weakening of the DBD-DOPC and DBD-DOPS hydrogen-bond structures should be attributed to the presence of cholesterol.
We have located three possible HB between DBD and cholesterol: for both H2 and H4 of DBD with hydroxyl’s oxygens of 
cholesterol as well as oxygens of DBD with hydroxyl’s hydrogen of cholesterol.  The distances are still typical between 1.7 
and 2.1~\AA, but as it can be seen in snapshot (D) of Fig.\ref{fig3}, none of them is due to double HB but to single ones. These
facts can be probably relevant regarding the interactions of drugs from the DBD family with cell membranes.

\subsubsection{Atomic site-site distances}

\begin{figure}[ht]
\begin{center}
\includegraphics[scale=0.4]{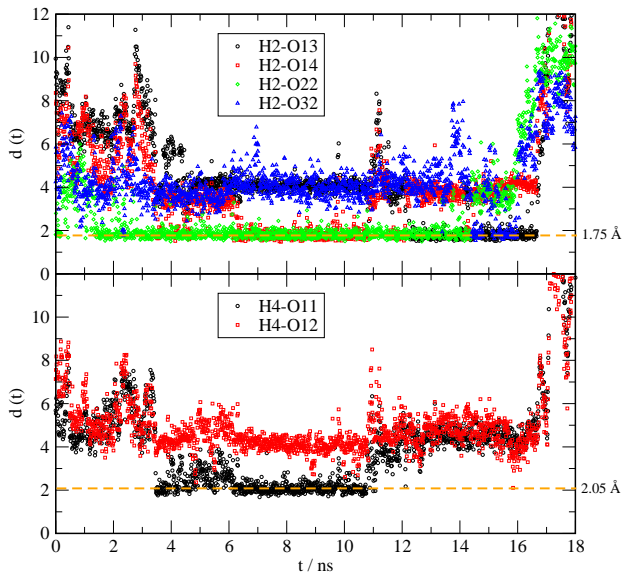}
\caption{\label{fig6}Distance distribution of selected sites in DBD-DMPC bonding (system 1) as a function of 
simulation time.  Top: Distances of DBD's H2 with DMPC's sites (O13, black circles; O14, red squares; O22, green diamonds
and O23, blue triangles). Bottom: Distances of DBD's H4 with DMPC's sites (O11, black circles; O12, red squares). 
Dashed lines indicate typical HB distances obtained from Fig.\ref{fig2} (1.75, 2.05~\AA$\,$). }
\end{center}
\end{figure}

\begin{figure}[ht]
\begin{center}
\includegraphics[scale=0.4]{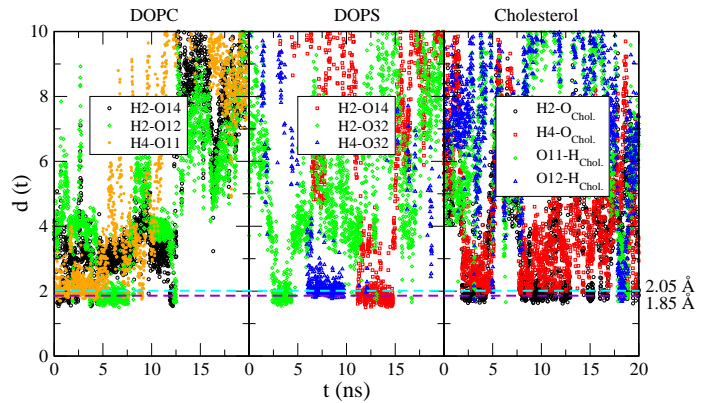}
\caption{\label{fig7}Distance distribution of selected sites in DBD-DOPC,  DBD-DOPS  and DBD-cholesterol (system 3)
as a function of simulation time.  Left: Distances of DBD-DOPC (H2-O14,  black circles; H2-O12,  green diamonds
and H4-O11, orange stars).  Middle: Distances of DBD-DOPS (H2-O14,  red squares; H2-O32,  green diamonds
and H4-O32, blue triangles).  Right: Distances of DBD-cholesterol (H2-O$_{chol.}$,  black circles; H4-O$_{chol.}$, 
red squares; O11-H$_{chol.}$,  green diamonds and O12-H$_{chol.}$,  blue triangles).  Dashed lines indicate  
typical HB distances obtained from Fig.\ref{fig5} (1.85 and 2.05~\AA$\,$). }
\end{center}
\end{figure}

After evaluating local structure of DBD, we have made a first step into the analysis of HB dynamics estimating the 
lifetime of some of the HB reported by RDF.  No other dynamical properties involving time-correlation functions such 
as power spectra\cite{marti1993computer,padro2004response},  relaxation times or self-diffusion coefficients\cite{padro1994molecular,marti2002microscopic} have been considered here.  In order to estimate 
the averaged time intervals for DBD-lipid association,  we display the time evolution of selected atom-atom distances 
$d(t)$ in Figures ~\ref{fig6} and ~\ref{fig7},  for systems 1 and 3, respectively.  Results for system 2 are in full 
agreement with those for system 3 and will be numerically reported in Table~\ref{tab2}.  We can see that typical 
hydrogen-bonding distances between 1.7-2.1~\AA$\,$ are reached in all cases.  For instance,  sites O13 and O14 
of DMPC are reported independently in order to provide information on the relative distances between DBD's 
H2 site and these two zwitterionic sites of DMPC.   Interestingly, we can observe that DBD's 'H4' is able to bind 
DMPC's 'O11' but not 'O12'.  Typical HB time lengths of about 5 ns are observed.  Conversely,  the other two 
selected distances are significantly longer (12 ns for H2 bound to DMPC's O22) or shorter (2 ns for H2 bound to O32) 
and correspond to limit boundaries.  The values reported in Fig.~\ref{fig6} are representative of the average values 
collected in Table~\ref{tab2} over full trajectories.  At the cholesterol-free membrane (system 1) we get typical 
HB lifetimes between 5 and 8 ns, whereas in systems 2 and 3 the range is broader, between 2 and 8 ns for lipids DOPC and 
DOPS and, interestingly, of only 1-2 ns for DBD-cholesterol hydrogen bonds.  These shorter HB are represented in 
Fig.~\ref{fig7}, where the HB dynamics indicates a larger extent of fluctuations, with bonds continuously formed and broken.  


When analysing the hydrogen-bonding of DBD with cholesterol (system 3), we have observed that some periods of hydrogen-bonding are established between oxygens 'O11'and 'O12' of DBD and the hydroxyl's hydrogen of cholesterol ('H$_{\rm Chol.}$').  In order to compute more precise hydrogen-bond lifetimes, correlation functions should be used (see Ref.\cite{marti2000dynamic} for instance),  but these are out of the scope of this paper.  A closer look indicates that in the case of DBD-cholesterol interactions, the longest living are the HB formed by hydrogens of a DBD (H2, H4 acting as donors) and the oxygens of cholesterol, the acceptors. ``Reverse'' hydrogen-bonding composed by cholesterol's hydrogen as the donors and oxygens O11-O12 of DBD (acceptors) is also possible but it is weaker than the former, with significantly smaller free-energy barriers as we will see below. 

\begin{table}
  \caption{\label{tab2} Distances between selected sites of DBD and the membrane.  Continuous time intervals are obtained from averaged computations.}
\fontsize{7}{10}\selectfont

\begin{tabular} {|c|c|c|c|c|c|}  \hline
DBD site& System & Membrane  & Lipid sites & Distance  (\AA)  & Lifetime (ns) \\
    \hline
     H2   & 1 &  DMPC & O13-O14   &  1.7 & 5  \\
     H2  &  1&  DMPC &  O22-O32   &  1.7 & 8    \\
     H4  &  1 & DMPC &  O11   &  1.7 &  5   \\   
     H4  &  1 & DMPC &  O12  &  4.0 &  5  \\     \hline
     H2   &  2 & DOPC & O13-O14   &  1.85 & 3  \\
     H2  &  2 & DOPC &  O22-O32   &  1.85 & 2.75    \\
     H2  &  2 & DOPC &  O11   &  4.0 &  6    \\
     H2  &  2 & DOPC &  O12   &  1.75 & 6  \\  
     H4   & 2 & DOPC & O11-O12   &  2.0  & 3 \\
     H2  &  2 & DOPS &  O13-O14 &  1. 7 & 2    \\
     H2  &  2 & DOPS &  O22-O32   &  1.7 &  8  \\  \hline 
     H2 &  3 & DOPC & O13-O14 &   1.85 & 2.5  \\
     H2  &  3 & DOPC &  O22-O32   &  1.85 & 4    \\
     H4  &  3 & DOPC &  O11  &  2.0 &  5    \\
     H2 &  3 & DOPS & O13-O14 &   1.85 & 2.5  \\
     H2  &  3 & DOPS &  O22-O32   &  1.85  & 2    \\
     H4   & 3 & DOPS & O22-O32   &  2.1  & 3 \\
     H2  &  3 & Cholesterol &  O$_{\rm Chol.}$ &  1. 85 & 2    \\
     H4  &  3 & Cholesterol &  O$_{\rm Chol.}$ &  2.1  & 2    \\
     O11-O12  & 3  & Cholesterol &  H$_{\rm Chol.}$ &  1.85  & 1    \\  
    \hline
\end{tabular}
\end{table}

\subsection{Potentials of mean force between DBD and lipids}
\label{pmfs}

The use of a variety of one-dimensional methods to compute the potentials of mean force (PMF) between two tagged species has been extensively discussed in the literature (see Ref.\cite{trzesniak2007comparison} for a review), where up to twelve methods were applied to the benchmark case of a methane pair in aqueous solution. The authors concluded that the best choice is a constraint-bias simulation combined with force averaging for Cartesian or internal degrees of freedom. The results from unbiased simulations, as those reported in the present 
work, were considered good at the qualitative level, with the PMF reasonably well reproduced.  However, the use of one-dimensional 
reaction coordinates is simply an approximation to the real ones\cite{geissler2001autoionization}, which may be in general multidimensional, presumably involving a limited number of water molecules and, eventually coordinates or distances to the other species of the system. Methods which do not assume any preconceived reaction coordinates such as transition path sampling\cite{marti2000stochastic,bolhuis2002transition,marti2004transition} or, those allowing to consider collective variables,  such as metadynamics\cite{barducci2008well,lu2020long} would be in order to obtain much more accurate free-energy landscapes but they require a huge amount of computational time.  Since the determination of reaction coordinates for the adsorption of DBD at zwitterionic membranes is out of the scope of this paper, we will consider the radial distances between two species as our order parameters useful to work as reaction coordinates of unbiased simulations. 

In such case, we can easily obtain a good approximation of the PMF through the so-called reversible work  $W_{AB}(r)$ required to 
move two tagged particles (A,B) from infinite separation to a relative separation $r$ (see for instance Ref.\cite{chandler1987introduction}, 
chapter 7):

\begin{equation}
W_{AB}(r) = -\frac{1}{\beta}\ln g_{AB}(r),
\label{eq2}
\end{equation}

where $\beta = 1/(k_B T)$ is the Boltzmann factor, $k_B$ the Boltzmann constant and T the temperature.  

\begin{figure}[ht]
\begin{center}
\includegraphics[scale=0.4]{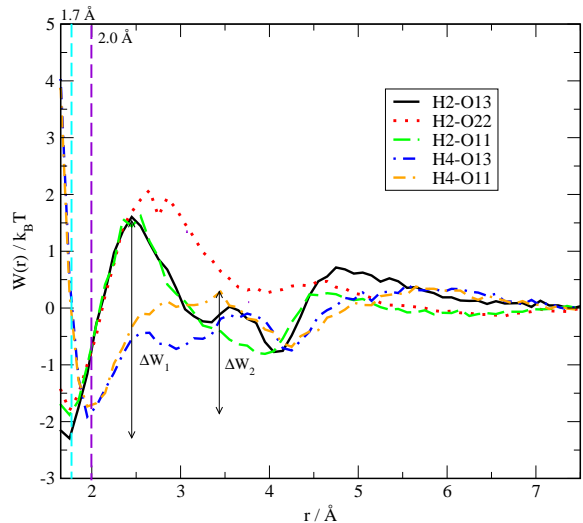}
\caption{\label{fig8}Potentials of mean force for the adsorption of  DBD to DMPC (system 1).}
\end{center}
\end{figure}

We show PMF for system 1 in Fig.~\ref{fig8}.  A free-energy barrier defined by a neat first minimum and a first maximum of $W(r)$ 
is clearly seen in all cases,  with size $\Delta W_1$ (H2 hydrogens of DBD) and $\Delta W_2$ (H4 hydrogens of DBD).  As expected the locations directly match the first maxima of the corresponding RDF.  At first sight we can observe that $\Delta W_1 > \Delta W_2$ in 
all cases, which is a clear indication that H2 bonding is significantly stronger than that of H4.  The full set of positions and 
free-energy barriers for the three systems has been reported in Table~\ref{tab3}.  There we can observe overall barriers between
1.3 and 4.6 $k_{B}T$, what correspond to 0.77-2.74 kcal/mol.  We observe stable binding distances very close to the typical hydrogen-bond distances. Furthermore, since the typical energy of water-water hydrogen-bonds estimated from {\it ab-initio} calculations is of about 5 kcal/mol\cite{feyereisen1996hydrogen}, we could conclude that our result are probably underestimated.

Given the absence of DBD free-energy barriers in the literature (up to our knowledge), and for the sake of comparison with other similar systems, the PMF of tryptophan in a di-oleoyl-phosphatidyl-choline bilayer membrane shows a barrier of the order of 4 kcal/mol\cite{maccallum2008distribution}, whereas the barrier for the movement of tryptophan attached to a poly-leucine $\alpha$-helix
inside a DPPC membrane was reported to be of 3 kcal/mol\cite{de2013role}. Finally, neurotransmitters such as glycine, acetylcholine or glutamate were reported to show small barriers of about 0.5-1.2 kcal/mol when located close to the lipid glycerol backbone\cite{peters2014interaction}. These values could further indicate that our estimations match at least the order of magnitude
of the free-energy barriers.

\begin{table}
  \caption{\label{tab3} Free-energy barriers $\Delta$W for the binding of DBD to water and lipids.  In order to quantify the height of all barriers,  1 $k_{B}T = 0.596$ kcal/mol.  Site '1' corresponds to DBD and 'Site 2' to water or a lipid.}
\fontsize{7}{10}\selectfont
\begin{tabular} {|c|c|c|c|c|c|}  \hline
System & Site 1  &  Site 2 &  r$_{min.}$ (\AA) &  r$_{max.}$  (\AA) & $\Delta W$ ($k_{B}T$)  \\    \hline
1 & H2    &   O13-O14 (DMPC)   &    1.75   &   2.46   &    3.9    \\
1 & H2    &   O22-O23 (DMPC)   &    1.75    &   2.67   &    3.8   \\  
1 & H2    &   O11 (DMPC)   &    1.75   &   2.46   &    3.4     \\
1 & H4    &   O13-O14 (DMPC)   &    2.0    &   3.6   &   1.8   \\  
1 & H4    &   O11 (DMPC)   &    2.0    &   3.44   &   2.0    \\    \hline
2 & H2   &  O  (Water)  &  1.85  & 2.55   &  2.5  \\
2 & H2  &  O13-O14 (DOPC)   &    1.75   &   2.6   &   4.1   \\
2 & H2    &   O22-O23 (DOPC)   &    1.8   &   2.7   &    3.8   \\
2 & H2    &   O11 (DOPC)   &    1.75   &   2.5   &    2.9    \\
2 & H4    &   O13-O14 (DOPC)   &    1.95    &   3.56   &   2.4    \\  
2 & H4    &   O11 (DMPC)   &    1.95    &   2.9   &   2.3    \\  
2 & H2  &  O13-O14 (DOPS)   &    1.75   &   2.55   &   4.6   \\
2 & H2    &   O22-O23 (DOPS)   &    1.75   &   2. 95  &    4.1   \\
2 & H2    &   O11 (DOPS)   &    2.0   &   2.8   &    1.3    \\
2 & H4    &   O13-O14 (DOPS)   &   2.0   &   3.56   &   1.7    \\   \hline
3 & H2   &  O  (Water)  &  1.85  & 2.55   &  2.5    \\
3 & H2  &  O13-O14 (DOPC)   &    1.75   &   2.47  &   3.8   \\
3 & H2    &   O22-O23 (DOPC)   &    1.75   &   2.57   &    3.9  \\
3 & H2    &   O11 (DOPC)   &    1.75   &   2.33   &    2.5   \\
3 & H4    &   O11 (DOPC)   &    1.97    &   2.76   &   1.7    \\  
3 & H2  &  O13-O14 (DOPS)   &    1.75   &   2.36   &   3.6   \\
3 & H2    &   O22-O23 (DOPS)   &    1.75   &   2. 45  &    4.0   \\
3 & H4    &   O22 (DOPS)   &   1.97   &   2.97   &   3.1    \\  
3 & H2   & O$_{\rm Chol.}$  &  1.87  & 2.86  &   3.5  \\
3 & H4  &  O$_{\rm Chol.}$  &  2.05  & 3.95  &   2.2  \\
3 & O11-O12 & H$_{\rm Chol.}$  &  1.87  & 2.94  &  1.65 \\       \hline
\end{tabular}
\end{table}

\section{Concluding remarks}
\label{concl}

A series of molecular dynamics simulations of a 3,4-dihydro-1,2,4-benzothiadiazine-1,1-dioxide molecule embedded in 
phospholipid bilayer membranes in aqueous ionic solution have been performed by molecular dynamics using the 
CHARMM36m force field. We have focused our analysis on the local structure of DBD,  when associated to lipids, water 
and cholesterol molecules.  After the systematic analysis of meaningful data, we noted relevant changes in local structure 
and dynamics of DBD.  The location of DBD in the interface of the membrane is permanent when the bilayer is formed 
only with DMPC lipids.  However, when DOPC and DOPS replace the DMPC backbone of the membrane,  DBD is able to 
make excursions to the solvent water. The same feature has been observed in the most realistic case,  when a membrane 
formed by 56$\%$ of DOPC, 14$\%$ of DOPS and 30$\%$ of cholesterol in sodium chloride solution has been set.

We have computed radial distribution functions defined for the most reactive particles, namely hydrogens 'H2' and 'H4'
and oxygens of DBD, according to labels defined in Figure~\ref{fig1} as well as selected sites of lipids and cholesterol
able to form hydrogen bonds with DBD.  Our data revealed the existence of a strong first coordination shell and a milder 
second coordination shell for DBD-lipid structures.  Such first shell is due to hydrogen bonds of variable lengths, between 
1.7 and 2.1~\AA$\,$, in good agreement with experimental data obtained from fluorescence measurements\cite{liu2013fluorescence} 
for similar small-molecule-membrane systems.  A direct analysis based on monitoring the relative distances between 
tagged sites of DBD and lipids has revealed that the lifetime of such hydrogen bonds ranges (obtained by averaging 
data from the 200 ns MD trajectories simulated) between 1 ns for DBD-cholesterol and up to 8 ns in the case of the 
HB formed between H2 of DBD and sites O22-O32 of DMPC and DOPC.
 
Our results indicate that most of stable bonds formed by DBD and lipids usually involve sites H2 and H4 of DBD 
in double bonding with selected sites, as indicated in Figure~\ref{fig3}.  Only in the case of DBD-cholesterol bonds these
are formed in a single-bond status, either between the hydrogen from the hydroxyl group of cholesterol and DBD's oxygens
or between the oxygen from the hydroxyl group of cholesterol and DBD's species H2 and H4.  Finally, from the analysis of
potentials of mean force based on reversible work calculations, we have estimated the free-energy barriers of the HB
reported above (see Table~\ref{tab3}), where the strongest corresponded to the association between DBD's H2 and 
oxygens sites O13-O14 of DOPS in the absence of cholesterol.  The influence of cholesterol has been especially noted
in the weakening of the DBD-lipid hydrogen bond connections and it may be a relevant factor when the interaction of
drugs from the DBD family is considered from a pharmaceutical perspective. 

\section*{Data availability}

The data that support the findings of this study are available from the corresponding author upon reasonable request.

\begin{acknowledgments}
The authors gratefully acknowledge financial support provided by the Spanish Ministry of Economy and Knowledge (grant 
PGC2018-099277-B-C21, funds MCIU/AEI/FEDER, UE).  ZH is the recipient of a grant from the China Scholarship Council 
(number 202006230070). 
\end{acknowledgments}

\newpage

\bibliography{references}


\end{document}